\documentclass[cameraready]{Interspeech}

\usepackage{amsmath,graphicx,amssymb,booktabs}
\usepackage{multirow}
\usepackage{hyperref}
\usepackage{xcolor}
\usepackage{parskip}
\graphicspath{{figures/}}
\AtBeginDocument{\pagestyle{plain}\setlength{\footskip}{18pt}}

\title{ACR-Net: Mitigating Semantic Dominance via Contrastive Acoustic-Semantic Decoupling}
\author[affiliation={1}, orcid=0009-0004-1250-5505, equalcontribution]{Mengke}{Zhang}
\author[affiliation={1}, orcid=0009-0003-9464-3620,equalcontribution]{Yanda}{Shao}
\author[affiliation={1}, orcid=0009-0006-0836-5575,equalcontribution]{Tianhe}{Wu}
\author[affiliation={1}, orcid=0009-0002-6232-517X,correspondingauthor]{Kai}{Feng}

\address{
    $^1$ Beijing University of Posts and Telecommunications, China
}
\email{zmk1460997475@bupt.edu.cn, 972359872@bupt.edu.cn, 2934221533@qq.com, fk050313@bupt.edu.cn}

\begin{document}

\maketitle
\thispagestyle{plain}
\hypersetup{
  pdftitle={ACR-Net: Mitigating Semantic Dominance via Contrastive Acoustic-Semantic Decoupling},
  pdfauthor={Mengke Zhang, Yanda Shao, Tianhe Wu, Kai Feng},
  pdfsubject={Accepted at Interspeech 2026}
}
\begin{abstract}

Speech Emotion Recognition (SER) models and Audio LLMs fail when vocal tone contradicts textual semantics. Extensive training on congruent data engenders a rigid semantic prior, causing ``Semantic Dominance'', a severe text bias during cross-modal conflicts. We introduce the ASPIRE benchmark of adversarial audio-text pairs across four conflict types, alongside two metrics: Semantic Overconfidence Penalty (SOP) and Latent Decoupling Degree (LDD). Unlike fusion methods that merely re-weight collapsed features, our Acoustic Conflict Resolution Network (ACR-Net) uses Cross-Modal Attention and Contrastive Decoupling Loss to disentangle contradictory representations into orthogonal latent spaces. Experiments show existing models fail under polarity conflicts due to extreme semantic overconfidence. Conversely, ACR-Net preserves acoustic fidelity, maintaining high LDD to drive SOP near zero, achieving superior Acoustic Accuracy (ACC) in both adversarial and standard scenarios.

\par
\noindent Code and dataset: \url{https://github.com/zmkshakespar/ACR-Net}.
\end{abstract}

\section{Introduction}

\begin{figure}[t]
\centering
\includegraphics[width=\linewidth]{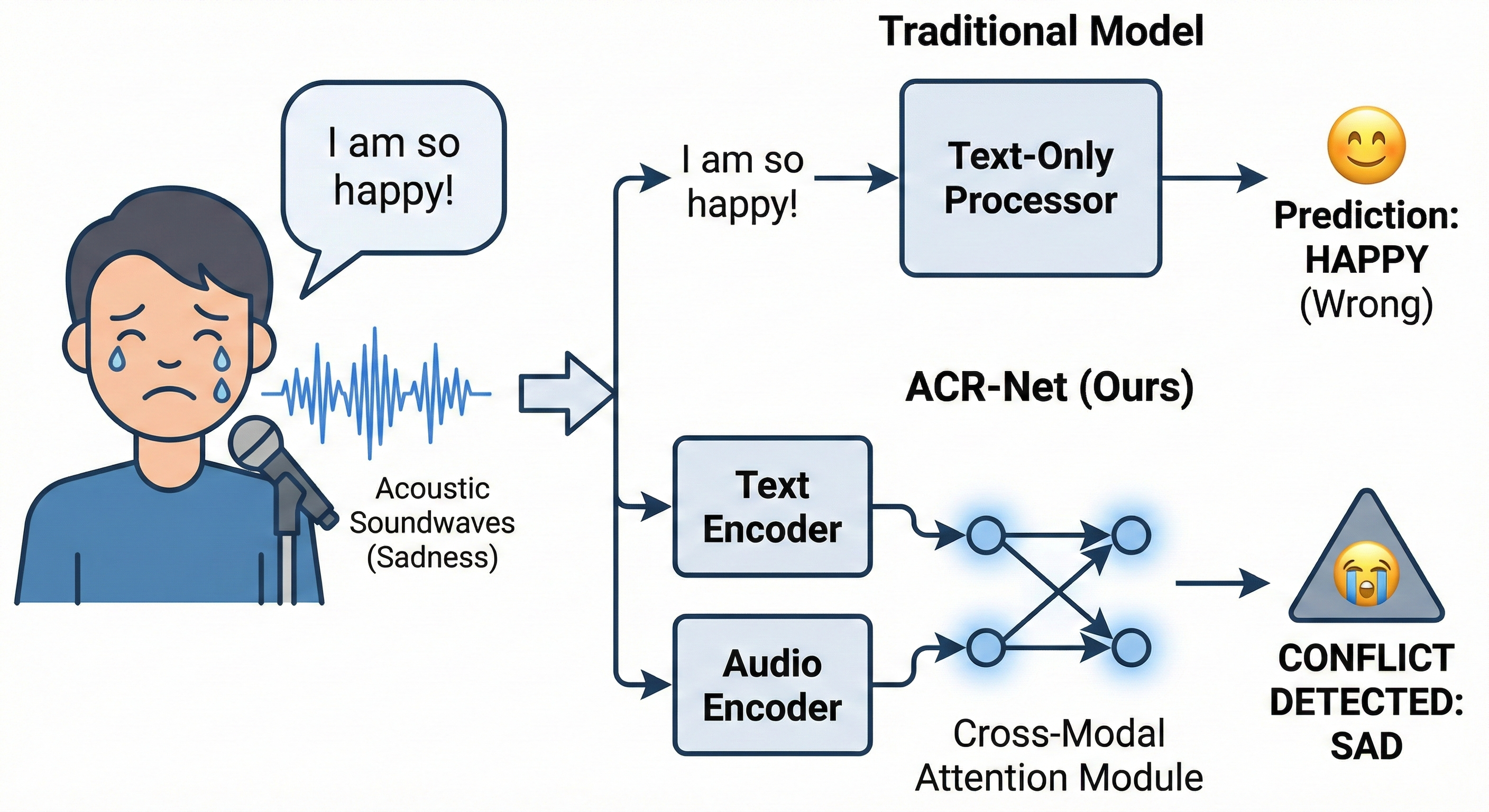}
\caption{An illustration of Semantic Dominance in emotional conflict. Traditional multi-modal models blindly trust the semantic content, whereas our approach detects the acoustic-semantic contradiction.}
\label{fig:problem}
\end{figure}

Speech Emotion Recognition (SER) is pivotal for human-computer interaction. While pure acoustic models like WavLM \cite{chen2022wavlm} and emotion2vec \cite{ma2024emotion2vec} capture rich paralinguistic features, multi-modal models (e.g., Whisper \cite{radford2023robust}) and Audio LLMs (e.g., Qwen-Audio \cite{chu2023qwen}) leverage linguistic context for SOTA results. However, these models inherently assume acoustic-semantic congruence, failing in ``Implicit Discrepancy'' scenarios (e.g., sarcasm) where vocal tone contradicts text (as illustrated in Figure \ref{fig:problem}).

Early efforts addressed this via specialized sarcasm detection (e.g., MUStARD \cite{castro2019towards}) or complex fusion routing. Yet, they succumb to \textit{modality collapse} \cite{wang2020what}, as textual tokens provide low-variance \textit{semantic shortcuts} \cite{bias_purification_eccv2024, du2024towards, geirhos2020shortcut} that overshadow subtle acoustic cues. Standard datasets (e.g., IEMOCAP \cite{busso2008iemocap, lotfian2017retrieving}, RAVDESS \cite{livingstone2018ravdess}) exacerbate this bias by filtering out incongruent samples to artificially inflate annotator agreement.

Heavily fine-tuned on such congruent data \cite{bitton2024data}, modern Audio LLMs develop rigid semantic priors. Probing studies (e.g., MCR-Bench \cite{wang2025mcrbench, correa2025emis}) confirm a catastrophic ``Semantic Dominance,'' causing models to disregard acoustic evidence during conflicts. Mitigations like the Multimodal Consensus Router (MCR) \cite{mcr2025router} attempt fusion-level weight calibration, but merely shift decision boundaries \cite{peng2022balanced} without disentangling underlying representations. This masked vulnerability necessitates controlled diagnostic benchmarks and specialized decoupling architectures.

To address this, we introduce the \textbf{ASPIRE} (Acoustic-Semantic Probing for Inconsistency REsolution) benchmark. Synthesized via the EmotiVoice engine \cite{emovoice_paper}, ASPIRE forces acoustic-semantic contradictions to stress-test reliability. To rigorously quantify cognitive blind spots, we propose two probing metrics: Semantic Overconfidence Penalty (SOP) and Latent Decoupling Degree (LDD). Evaluations reveal a stark reality: under implicit discrepancy, modern multi-modal models exhibit extreme semantic overconfidence (high SOP) and heavily entangled feature spaces (low LDD).

To overcome this structural vulnerability, we propose the Acoustic Conflict Resolution Network (\textbf{ACR-Net}). Utilizing a structurally isolated dual-stream architecture with a frozen Whisper backbone and Low-Rank Adaptation \cite{hu2021lora}, ACR-Net employs Cross-Modal Attention (CMA) \cite{tsai2019multimodal} to actively scan for audio-text inconsistencies. Crucially, we introduce a Contrastive Decoupling Loss to explicitly separate conflicting representations in the latent space, transcending simple weight-shifting to achieve true feature decoupling.

\textbf{Concurrent Work.} The concurrent FAS framework and CASE dataset \cite{concurrent2026when} also explore acoustic-semantic conflict. Our work fundamentally diverges: first, instead of naturalistic collections with ambiguous emotional boundaries, ASPIRE adopts an \textit{adversarial synthesis} paradigm ensuring absolute variable isolation. Second, whereas FAS focuses on fusion-level mitigation, ACR-Net addresses the root cause by explicitly decoupling representations to overcome rigid semantic priors.

Ultimately, we present the first systematic approach to resolving implicit acoustic-semantic discrepancy via explicit feature decoupling. Our contributions are:
\begin{enumerate}
    \item \textbf{ASPIRE Benchmark}: An adversarial diagnostic dataset of four structural conflict types, enabling strict evaluation of semantic dominance without data contamination risks.
    \item \textbf{Diagnostic Metrics}: SOP and LDD to rigorously quantify modality bias and latent feature entanglement prior to fusion.
    \item \textbf{ACR-Net}: A dual-stream architecture that explicitly disentangles conflicting representations via a Contrastive Decoupling Loss, achieving state-of-the-art Acoustic Accuracy (ACC) under severe cross-modal interference.
\end{enumerate}

\section{The ASPIRE Benchmark}
\label{sec:benchmark}

\subsection{Dataset}
While recent benchmarks like MCR-Bench \cite{wang2025mcrbench} and the CASE dataset \cite{concurrent2026when} have pioneered the exploration of cross-modal conflict, they primarily rely on manually collected data. The emotional orientation in such data is heavily influenced by \textbf{individual subjective factors} and ambiguous boundaries, making it difficult to maintain strict variable isolation. To address this, we introduce the \textbf{ASPIRE} (Acoustic-Semantic Probing for Inconsistency REsolution) benchmark. We synthesized ASPIRE using a rigorous pipeline aimed at generating unambiguous samples. We utilized the 1.5B parameter version of Emotivoice \cite{emovoice_paper}, a highly controllable \textbf{TTS model} that supports \textbf{fine-grained emotional control} (e.g., expressing ``happiness with a hint of grievance'').

\subsubsection{Synthesis Pipeline}
Finding high-quality, perfectly isolated conflict data in the real world is nearly impossible. Thus, we adopted a four-step adversarial synthesis pipeline:

\begin{enumerate}
    \item \textbf{Prompt Generation:} We designed 1,800 JSONL prompts pairing contradictory and congruent audio-text instructions. Emotivoice generated 1,523 raw WAV files, immediately discarding 277 severe generation failures.
    \item \textbf{Standardization:} Files were resampled from 22kHz to 16kHz using FFmpeg. A size filter removed corrupted clips under 5KB, leaving 1,495 valid samples.
    \item \textbf{Automated Acoustic Filtering:} We employed emotion2vec+ \cite{ma2024emotion2vec} for automated acoustic validation. To ensure distinct paralinguistic cues, we strictly discarded samples where the prediction confidence was below 50\% or mismatched the intended acoustic prompt, yielding 1,310 candidates.
    \item \textbf{Expert Auditory Review:} Human experts performed a final auditory screening. They approved only natural-sounding samples where the intended emotional conflict was immediately perceptible, removing robotic or ambiguous outputs.
\end{enumerate}

\subsubsection{Dataset Structure}
After the pipeline, 1,200 high-quality samples were selected (800 conflict, 400 normal). To provide a standardized protocol, the samples are strictly partitioned using 5-fold cross-validation (960 train/fine-tune, \textbf{240 held out for validation} per fold). Furthermore, we \textbf{extra-synthesized an independent test set of 200 samples} to facilitate benchmarking.

The conflict data encompasses 19 distinct adversarial pairings (out of 24 possible incongruent combinations, 5 extreme cases were discarded due to persistent TTS failure), formed by crossing 4 target acoustic emotions (Neutral, Sad, Surprised, Angry) with 7 textual emotions (Angry, Sad, Happy, Surprised, Neutral, Tender, Fearful). Based on generation retention rates and acoustic properties, we categorize these conflicts into four structural types, detailed in Table \ref{tab:dataset_structure}.
\begin{table}[t]
  \caption{\textbf{Structural Categorization of Conflict Pairings in ASPIRE.}}
  \vspace{-2mm}
  \label{tab:dataset_structure}
  \centering
  \renewcommand{\arraystretch}{1.4} 
  \resizebox{\columnwidth}{!}{%
  \begin{tabular}{@{} p{3.0cm} c p{4.6cm} @{}}
    \toprule
    \textbf{Conflict Type} & \textbf{Retention} & \textbf{Characteristics \&} \\
    \textit{(Example)} & \textbf{Rate} & \textbf{Synthesis Difficulty} \\
    \midrule
    \textbf{Polarity Opposite} \newline \textit{\small (Sad + Happy)} & 45\% & \textbf{Hardest.} Extreme polarity clash; text semantics heavily overpower acoustic prompts. \\
    \cmidrule(lr){1-3} 
    \textbf{Arousal Opposite} \newline \textit{\small (Neutral + Angry)} & 84\% & \textbf{Easiest.} Neutral acoustics are stable, but high-arousal acoustics easily collapse. \\
    \cmidrule(lr){1-3}
    \textbf{Valence Opposite} \newline \textit{\small (Sad + Angry)} & 79\% & \textbf{Mixed.} Distinct emotional modalities forming genuine acoustic-semantic blending. \\
    \cmidrule(lr){1-3}
    \textbf{Emotion Masking} \newline \textit{\small (Neutral + Sad)} & 68\% & \textbf{Bipolar.} Simulates suppression. Easy for low-arousal text, hard to mask positive text. \\
    \bottomrule
  \end{tabular}%
  }
\end{table}
This detailed breakdown reveals that TTS models inherently suffer from a form of semantic dominance during generation, especially in Polarity Opposite pairs where textual sentiment overwrites vocal instructions. 

\subsection{Evaluation Metrics}
Relying solely on classification accuracy is insufficient to diagnose the root causes of failure in conflict scenarios. To rigorously quantify a model's cognitive blind spots and its internal representation quality, we propose two novel, specialized metrics:

\textbf{1. Semantic Overconfidence Penalty (SOP):}
When models fail, they often do not fail randomly; they fail by blindly trusting the text. SOP quantifies this "blindness" by measuring the margin of probability advantage the text modality holds over the acoustic modality during prediction. For a dataset of $N$ conflict samples:
\begin{equation}
    \mathrm{SOP} = \frac{1}{N} \sum_{i=1}^{N} \max\Big(0, P(\hat{y}_{\text{text}}^{(i)}) - P(\hat{y}_{\text{audio}}^{(i)})\Big)
\end{equation}
where $P(\hat{y}_{\text{text}}^{(i)})$ and $P(\hat{y}_{\text{audio}}^{(i)})$ denote the predicted Softmax probabilities assigned to the text's apparent emotion and the audio's true emotion, respectively. A higher SOP indicates severe semantic dominance, whereas a lower SOP suggests the model exhibits ``hesitation'' and resists absolute textual deception.

\textbf{2. Latent Decoupling Degree (LDD):}
To verify whether a model fundamentally resolves conflict rather than merely shifting weights, we must probe its feature space prior to final fusion. LDD measures the orthogonality of the pooled acoustic representation ($\mathbf{h}_a$) and semantic representation ($\mathbf{h}_t$) in the latent space:

\begin{equation}
    \mathrm{LDD} = \frac{1}{N} \sum_{i=1}^{N} \left( 1 - \frac{\mathbf{h}_a^{(i)} \cdot \mathbf{h}_t^{(i)}}{\lVert \mathbf{h}_a^{(i)} \rVert_2 \lVert \mathbf{h}_t^{(i)} \rVert_2} \right)
\end{equation}

A higher LDD mathematically proves that the model explicitly disentangles contradictory features, pushing them apart into independent sub-spaces, which is essential for overcoming rigid semantic priors.

\section{Methodology}
\label{sec:method}

To overcome Semantic Dominance, we propose the Acoustic Conflict Resolution Network (\textbf{ACR-Net}). Unlike traditional multi-modal approaches that seek a compromised, unified representation through fusion, ACR-Net is designed around the principle of \textbf{Explicit Feature Decoupling}. By forcing acoustic and semantic features into orthogonal sub-spaces, we prevent the dominant linguistic priors from overwriting subtle vocal emotions.

\subsection{Dual-Stream Modality Encoders}
To establish independent feature spaces before conflict resolution, ACR-Net employs a structurally isolated dual-stream architecture.

\textbf{1. Audio Stream (Acoustic Preservation):} We utilize the encoder of \textbf{Whisper-Medium} \cite{radford2023robust} to process the audio signal. Crucially, to prevent the pre-trained model from falling back on its text-heavy latent bias, we freeze the original weights and inject \textbf{LoRA} (Low-Rank Adaptation) \cite{hu2021lora} layers into every self-attention block. This allows the audio stream to specifically learn highly varying, fine-grained acoustic cues (e.g., breathiness, pitch shifts) independently of the text. The output is a sequence of acoustic embeddings $E_a \in \mathbb{R}^{T_a \times d}$.

\textbf{2. Text Stream (Semantic Anchor):} The transcript is processed by the frozen text encoder from Whisper. By strictly freezing these parameters, we preserve its robust linguistic understanding, extracting a pure semantic sequence $E_t \in \mathbb{R}^{T_t \times d}$.

\subsection{Cross-Modal Attention as Conflict Detector}
Instead of using attention to "blend" modalities, we repurpose the Cross-Modal Attention (CMA) \cite{tsai2019multimodal} block as an \textit{active inconsistency detector}. 

We map the acoustic embeddings $E_a$ to the Query matrix ($Q$), and the semantic embeddings $E_t$ to the Key ($K$) and Value ($V$) matrices:
\begin{equation}
    CMA(E_a, E_t) = \text{Softmax} \left( \frac{Q_a K_t^\top}{\sqrt{d}} \right) V_t
\end{equation}
In this formulation, the acoustic features actively "scan" the semantic sequence to locate the exact linguistic tokens that contradict the vocal tone. By applying a residual connection and mean pooling over the time dimension, we obtain temporally aligned representations ready for spatial decoupling.

\subsection{Contrastive Decoupling Loss}
The core innovation of ACR-Net lies in its optimization objective. Traditional cross-entropy loss implicitly encourages feature entanglement and lacks explicit margin constraints \cite{khosla2020supervised}. To break the network's reliance on text, we introduce a hybrid loss function:
\begin{equation}
    \mathcal{L}_{total} = \mathcal{L}_{CE} + \lambda \mathcal{L}_{Decouple}
\end{equation}

While $\mathcal{L}_{CE}$ trains the linear classifier to predict the correct acoustic emotion, $\mathcal{L}_{Decouple}$ acts as a spatial separator. We calculate the cosine similarity $S_i = \cos(\mathbf{h}_a^{(i)}, \mathbf{h}_t^{(i)})$ between the pooled acoustic vector ($\mathbf{h}_a$) and semantic vector ($\mathbf{h}_t$). The contrastive margin loss is defined as:
\begin{equation}
    \mathcal{L}_{Decouple} = \frac{1}{N} \sum_{i=1}^{N} \Big[ y_i(1 - S_i) + (1-y_i) \max(0, S_i - m) \Big]
\end{equation}
where $y_i = 1$ for normal samples and $y_i = 0$ for conflict samples. 
For implicitly discrepant samples, this loss explicitly penalizes similarity, forcing the contradictory audio and text representations apart by a margin $m$. This mathematical repulsion ensures that textual semantics cannot mask acoustic features, achieving true modality decoupling. The effectiveness of this decoupling strategy and its hyperparameter sensitivity are quantitatively validated in Table \ref{tab:strategy_comparison}.

\begin{table}[ht]
  \caption{\textbf{Comparison of Representation Strategies on the ASPIRE Benchmark.} ACR-Net explicitly disentangles features, achieving the highest ACC while minimizing SOP. The hyperparameter analysis shows the necessity of balanced decoupling.}
  \vspace{-2mm}
  \label{tab:strategy_comparison}
  \centering
  \renewcommand{\arraystretch}{1.2} 
  \resizebox{\columnwidth}{!}{%
  \begin{tabular}{l | c | c | c}
    \toprule
    \textbf{Representation Strategy} & \textbf{ACC (\%)} $\uparrow$ & \textbf{SOP} $\downarrow$ & \textbf{LDD} $\uparrow$ \\
    \midrule
    Naive Concatenation & 41.2 & 0.824 & 0.108 \\
    \midrule
    \multicolumn{4}{l}{\textit{Advanced Fusion Mechanisms}} \\
    + Tensor Fusion \cite{zadeh2017tensor} & 43.8 & 0.776 & 0.115 \\
    + MCR (Re-weighting) \cite{mcr2025router} & 48.5 & 0.628 & 0.242 \\
    \midrule
    \multicolumn{4}{l}{\textit{Explicit Decoupling Interventions (Ablation)}} \\
    + CMA (w/o Decouple Loss) & 53.0 & 0.534 & 0.334 \\
    + Decouple Loss (w/o CMA) & 62.1 & 0.253 & 0.697 \\
    \midrule
    \multicolumn{4}{l}{\textit{Hyperparameter Sensitivity (ACR-Net)}} \\
    \hspace{3mm} $\lambda=0.01, m=0.5$ (Weak Decoupling) & 60.4 & 0.312 & 0.510 \\
    \hspace{3mm} $\lambda=1.00, m=0.5$ (Over-Penalization) & 71.2 & 0.185 & \textbf{0.890} \\
    \hspace{3mm} $\lambda=0.10, m=0.1$ (Small Margin) & 64.8 & 0.254 & 0.620 \\
    \hspace{3mm} $\lambda=0.10, m=0.9$ (Excessive Margin) & 69.5 & 0.201 & 0.885 \\
    \midrule
    \textbf{+ ACR-Net (Optimal: $\lambda=0.1, m=0.5$)} & \textbf{76.5} & \textbf{0.128} & 0.864 \\
    \bottomrule
  \end{tabular}%
  }
\end{table}

\section{Experiments}
\label{sec:experiments}
\begin{table*}[t]
  \caption{\textbf{Acoustic Accuracy (ACC \%) comparison on standard congruent datasets.}}
  \vspace{-2mm}
  \label{tab:normal_results}
  \centering
  \renewcommand{\arraystretch}{1.15}
  \resizebox{0.9\textwidth}{!}{%
  \begin{tabular}{ll | c | c | c | c | c}
    \toprule
    \textbf{Model} & \textbf{Paradigm / Backbone} & \textbf{EmoDB}\cite{burkhardt2005emodb} & \textbf{CASIA}\cite{bin2008design} & \textbf{SAVEE}\cite{haq2008savee} & \textbf{CREMA-D}\cite{cao2014crema} & \textbf{RAVDESS}\cite{livingstone2018ravdess} \\
    \midrule
    \multicolumn{7}{l}{\textit{Category I: Audio-only SSL (Stable Acoustic Priors)}} \\
    \textcolor[HTML]{1F77B4}{\rule{1.5mm}{1.5mm}} \textbf{emotion2vec+ (2024)} & WavLM-base & \textbf{93.5} & \underline{89.5} & 85.4 & 74.2 & 89.2 \\
    \textcolor[HTML]{1F77B4}{\rule{1.5mm}{1.5mm}} \textbf{WavLM-Large (2022)} & WavLM-Large & \underline{91.2} & 85.5 & \underline{87.2} & \textbf{76.5} & 91.5 \\
    \midrule
    \multicolumn{7}{l}{\textit{Category II: Multi-modal Task Models (Semantically Anchored)}} \\
    \textcolor[HTML]{FF7F0E}{\rule{1.5mm}{1.5mm}} \textbf{Whisper-large-v3 (2023)} & Whisper-Enc (1.5B) & 76.8 & 73.2 & 82.5 & 71.4 & 92.4 \\
    \textcolor[HTML]{FF7F0E}{\rule{1.5mm}{1.5mm}} \textbf{SeamlessM4T (2023)} & Conformer (2.3B) & 75.2 & 71.8 & 80.4 & 69.8 & 87.2 \\
    \midrule
    \multicolumn{7}{l}{\textit{Category III: Audio LLMs (High-Prior / Prone to Data Leakage)}} \\
    \textcolor[HTML]{2CA02C}{\rule{1.5mm}{1.5mm}} \textbf{SALMONN (2024)} & Whisper+Vicuna (14B)& 79.2 & 81.4 & 84.1 & 72.5 & 89.0 \\
    \textcolor[HTML]{2CA02C}{\rule{1.5mm}{1.5mm}} \textbf{Qwen2-Audio (2024)} & Qwen-LM (8B) & 81.5 & \textbf{91.8} & 84.5 & 74.5 & \textbf{94.2} \\
    \midrule
    \multicolumn{7}{l}{\textit{Category IV: Recent Modal Interventions \& Ours (Conflict Resolution Focus)}} \\
    \textcolor[HTML]{D62728}{\rule{1.5mm}{1.5mm}} \textbf{MCR (2025)} & Re-weighting Fusion & 80.4 & 82.6 & 81.4 & 73.1 & 91.8 \\
    \textcolor[HTML]{D62728}{\rule{1.5mm}{1.5mm}} \textbf{FAS (2026)} & Adaptive Fusion & 83.2 & 84.5 & 83.8 & 74.5 & 92.0 \\
    \textbf{ACR-Net (Ours)} & \textbf{Decoupling (Whisper)} & 88.5 & 86.8 & \textbf{88.6} & \underline{75.8} & \underline{93.1} \\
    \bottomrule
  \end{tabular}%
  }
\end{table*}

\subsection{Experimental Setup}
We benchmark ACR-Net against diverse baselines, mapping corpora labels to four core classes (Angry, Happy, Sad, Neutral) for unified evaluation. To isolate architectural biases, baselines are grouped into four distinct categories:
\begin{itemize}
    \item \textcolor[HTML]{1F77B4}{\rule{1.5mm}{1.5mm}} \textbf{Audio-only SSL}: Models trained purely on acoustic features (emotion2vec+, WavLM-Large).
    \item \textcolor[HTML]{FF7F0E}{\rule{1.5mm}{1.5mm}} \textbf{Multi-modal Task Models}: Models designed for general speech tasks (Whisper-large-v3, SeamlessM4T).
    \item \textcolor[HTML]{2CA02C}{\rule{1.5mm}{1.5mm}} \textbf{Audio LLMs}: Large language models with native audio capabilities (SALMONN, Qwen2-Audio (8B)).
    \item \textcolor[HTML]{D62728}{\rule{1.5mm}{1.5mm}} \textbf{Recent Modal Interventions}: Frameworks explicitly addressing multimodal alignment and fusion calibration (MCR, FAS).
\end{itemize}

All baseline models are evaluated zero-shot to ensure rigorous testing. 

\subsection{Results on Normal Scenes}
To verify that ACR-Net maintains representation quality on standard congruent speech, we select five datasets assessing cross-lingual robustness and potential data-leakage: \textbf{EmoDB} (German), \textbf{CASIA} (Chinese), \textbf{SAVEE} (British English), \textbf{CREMA-D} (English), and \textbf{RAVDESS} (North American English). 

The results in Table \ref{tab:normal_results} reveal a clear domain gap and contamination bias in foundation models. While Qwen2-Audio achieves inflated scores on RAVDESS and CASIA (likely due to instruction-tuning exposure), its performance drops significantly on the cross-lingual EmoDB zero-shot test. Conversely, ACR-Net maintains robust, competitive performance across all benchmarks, proving that contrastive decoupling successfully preserves foundational acoustic recognition capabilities.

\subsection{Deep Diagnostics on Conflict Scenes (ASPIRE)}
To rigorously diagnose vulnerabilities under implicit discrepancy, we evaluate state-of-the-art fusion interventions (MCR and FAS) against ACR-Net across the four structural conflict types defined in Section \ref{sec:benchmark}. 

Table \ref{tab:conflict_results} highlights the fundamental flaws of traditional fusion mechanisms against adversarial semantics. We summarize our core findings below:

\textbf{Findings 1. The Failure of Weight Calibration under Polarity Conflict}: Polarity Opposite represents extreme semantic dominance. As text tokens provide low-variance semantic shortcuts, linguistic features heavily overpower acoustic cues during extraction. While fusion-based models (MCR, FAS) attempt feature re-weighting, mathematically scaling an already collapsed acoustic vector cannot recover lost information. Consequently, they exhibit severe Semantic Overconfidence (SOP $> 0.5$) and fail catastrophically (ACC $< 45\%$).

\textbf{Findings 2. Feature Decoupling is Essential for Modality Preservation}: Across all four conflict type, ACR-Net maintains an LDD consistently above 0.8. By enforcing mathematical repulsion in the latent space via the Contrastive Decoupling Loss, ACR-Net prevents dominant semantic priors from overwriting acoustic representations. This structural isolation enables the Cross-Modal Attention to accurately localize inconsistencies, driving SOP near zero and securing state-of-the-art Acoustic Accuracy.

\textbf{Findings 3. Asymmetric Vulnerability in Arousal}: Arousal Opposite proves easiest for all models, as low-arousal text (e.g., neutral transcripts) exerts a weaker semantic gravitational pull, allowing fusion models to partially leverage strong acoustic cues. However, when simulating emotional suppression ("Emotion Masking"), fusion models regress significantly, whereas ACR-Net maintains robust diagnostic accuracy.
\begin{table}[t]
  \caption{\textbf{Diagnostic Evaluation on the ASPIRE Benchmark.} The table details the performance across four distinct structural conflict types. By maintaining a high Latent Decoupling Degree (LDD), ACR-Net effectively suppresses Semantic Overconfidence (SOP) and achieves state-of-the-art Acoustic Accuracy (ACC).}
  \vspace{-2mm}
  \label{tab:conflict_results}
  \centering
  \footnotesize 
  \renewcommand{\arraystretch}{1.15}
  \resizebox{\columnwidth}{!}{%
  \begin{tabular}{@{}l | l | c | c | c@{}}
    \toprule
    \textbf{Conflict Type} & \textbf{Model} & \textbf{ACC} $\uparrow$ & \textbf{SOP} $\downarrow$ & \textbf{LDD} $\uparrow$ \\
    \midrule
    \multirow{3}{*}{\begin{tabular}[c]{@{}l@{}}\textbf{Polarity Opposite}\\ \textit{\scriptsize (Sad + Happy)}\end{tabular}} 
    & MCR & 35.2 & 0.612 & 0.205 \\
    & FAS & 42.5 & 0.548 & 0.288 \\
    & \textbf{ACR-Net (Ours)} & \textbf{68.4} & \textbf{0.185} & \textbf{0.812} \\
    \midrule
    \multirow{3}{*}{\begin{tabular}[c]{@{}l@{}}\textbf{Arousal Opposite}\\ \textit{\scriptsize (Neutral + Angry)}\end{tabular}} 
    & MCR & 65.8 & 0.315 & 0.322 \\
    & FAS & 68.4 & 0.280 & 0.395 \\
    & \textbf{ACR-Net (Ours)} & \textbf{85.2} & \textbf{0.092} & \textbf{0.885} \\
    \midrule
    \multirow{3}{*}{\begin{tabular}[c]{@{}l@{}}\textbf{Valence Opposite}\\ \textit{\scriptsize (Sad + Angry)}\end{tabular}} 
    & MCR & 50.4 & 0.485 & 0.245 \\
    & FAS & 55.6 & 0.412 & 0.310 \\
    & \textbf{ACR-Net (Ours)} & \textbf{75.8} & \textbf{0.145} & \textbf{0.840} \\
    \midrule
    \multirow{3}{*}{\begin{tabular}[c]{@{}l@{}}\textbf{Emotion Masking}\\ \textit{\scriptsize (Neutral + Sad)}\end{tabular}} 
    & MCR & 45.2 & 0.520 & 0.228 \\
    & FAS & 52.8 & 0.465 & 0.295 \\
    & \textbf{ACR-Net (Ours)} & \textbf{78.5} & \textbf{0.118} & \textbf{0.865} \\
    \bottomrule
  \end{tabular}%
  }
\end{table}
\section{Conclusion}
This paper mitigates ``Semantic Dominance'' in Audio LLMs. We systematically expose this vulnerability via the ASPIRE benchmark and two novel diagnostic metrics: SOP and LDD. Evaluations reveal traditional fusion catastrophically collapses under polarity conflicts as dominant text overwrites subtle paralinguistics. To resolve this, ACR-Net employs a Contrastive Decoupling Loss to explicitly project contradictory features into orthogonal latent spaces. By structurally preserving acoustic fidelity, ACR-Net neutralizes semantic overconfidence, achieving state-of-the-art Acoustic Accuracy (ACC) across adversarial and congruent datasets. Ultimately, this work provides a rigorous architectural blueprint for next-generation foundational speech models \cite{rubenstein2023audiopalm} to genuinely interpret complex, indirect human intents like irony and emotion suppression.
\section{Acknowledgments}
The authors would like to express their deepest gratitude to their parents for their unwavering moral support, endless patience, and continuous encouragement throughout the course of this research.

\section{Generative AI Use Disclosure}
During the preparation of this manuscript, generative AI tools were utilized to assist the authors in structural refinement, language polishing, and syntax optimization to ensure clear and professional academic expression. The authors carefully reviewed and edited all outputs generated by the AI to align with the intended scientific meaning. The authors take full accountability for the final content, experimental results, and overall accuracy of this paper.
\bibliographystyle{IEEEtran}
\bibliography{references}

\end{document}